\documentstyle[aps,prb,twocolumn,epsf]{revtex}

\newcommand{\ie}{{\it i.e.\ }}
\newcommand{\etal}{\emph{et al.}}
\newcommand{\D}{\mathrm{d}\mathit{}}
\newcommand{\statdef}{n^{\text{static}}_{\text{d}}}
\newcommand{\defden}{n_{\text{d}}}
\epsfclipon

\begin{document}

\twocolumn[\hsize\textwidth\columnwidth\hsize
\csname @twocolumnfalse\endcsname


\title{Vortex dynamics in two-dimensional systems at high driving forces}
\author{Hans Fangohr$^{*\dag}$, Simon J. Cox$^{*}$, Peter A. J. de Groot$^{\dag}$,}
\address{$^{*}$Department of Electronics and Computer Science, \\
$^{\dag}$Department of Physics and Astronomy, \\
University of Southampton, Southampton, SO17 1BJ, United Kingdom} 
\date{\today}

\maketitle

\begin{abstract}
  We study numerically the dynamics of two-dimensional vortex systems
  at zero temperature. In addition to pinned states and turbulent
  plastic flow, we find motion of vortices in rough channels along the
  direction of the driving force. In this decoupled channel r\'{e}gime
  we demonstrate how topological defects mediate the phase slip of
  different channels moving with different velocities. We thus provide
  important confirmation of recent analytical work describing vortex
  dynamics at high driving forces such as the moving glass theory of
  Giamarchi and Le Doussal. For the largest driving forces we find
  that the channels couple and observe elastic motion.

\end{abstract}

]
\narrowtext

\tableofcontents

\section{Introduction}
 
Vortex dynamics in the presence of disordering pinning show a variety
of non-equilibrium physics and dynamic phase transitions.
Experiments,\cite{Thorel73,Bhattacharya93,Yaron94,Higgins96,Hellerqvist96,Marchevsky97,Pardo98}
numerical,\cite{Shi91,Moon96,Ryu96,Spencer97,Olson98} and
analytical\cite{Koshelev94,Giamarchi96,Balents97,Balents98,Scheidl98b,LeDoussal98} work suggest that a disordered static system of vortices shows
ordering at higher driving forces.  Koshelev and
Vinokur\cite{Koshelev94} predicted a dynamic phase transition between
plastic sliding for driving forces just above de-pinning and coherent
motion of crystalline structures at high driving forces. Subsequently,
Giamarchi and Le\ Doussal\cite{Giamarchi96} predicted that the
strongly driven and reordered system would be a Moving Glass, where
vortices move 
elastically-coupled 
along static channels, such that they flow in the
direction of the driving force along well-defined, nearly parallel
paths in the pinning potential. These optimal channels (in two
dimensions) or sheets (in three dimensions) show a roughness and are
predicted to be a static and reproducible feature of the disorder
configuration. 

Balents, Marchetti and Radzihovsky argued\cite{Balents97} that in
addition to elastically-coupled channels (no topological defects in the
system) at intermediate velocities a transverse moving
smectic\cite{Balents97,Balents98} would exist in which motion of
vortices in different channels is decoupled (topological defects
between the channels). Later
work\cite{Balents98,Scheidl98b,LeDoussal98} mainly supported the
initial findings of Giamarchi and Le Doussal with the addition of the
moving smectic as predicted by Balents, Marchetti and
Radzihovsky.\cite{Balents97}
Different names are in common use: the Moving Transverse
Glass\cite{LeDoussal98} (MTG), moving smectic\cite{Balents98} and
decoupled channels\cite{Scheidl98b} refer to the decoupled channel
motion, and the Moving Bragg Glass\cite{LeDoussal98} (MBG), moving
lattice\cite{Balents98} and coherent phase\cite{Scheidl98b} refer to
the r\'{e}gime of elastically coupled channels.

The theoretical
descriptions\cite{Giamarchi96,Balents97,Balents98,Scheidl98b,LeDoussal98} of
these dynamic phases are based on elastic theory and assume either the
absence (for the MBG) or the irrelevance (for the MTG) of topological
defects. In fact, the theory of Giamarchi and Le
Doussal\cite{Giamarchi96,LeDoussal98} describes both r\'{e}gimes with
the same equation, which is (nearly) exact for the MBG and remains an
effective description for the MTG. In this work we investigate, for
the first time, the role of topological defects in the MTG to check
the validity of assumptions entering the theory of Giamarchi and Le\ 
Doussal, and find them to be justified.

We review the dynamic phase diagram for a 2d vortex system in the
presence of random disorder varying on a length scale much smaller
than the vortex-vortex spacing. Our model describes rigid vortices in
thin-films or decoupled pancake vortices in layered materials. We
employ a modified cut-off\cite{Fangohr00c} to the appropriate
interaction force which corresponds to a logarithmic vortex-vortex
interaction potential.\cite{Clem91} After annealing a vortex system to
zero temperature, we apply an increasing driving force and study the
dynamics of the system systematically for different pinning strengths.

Sec. \ref{sectionthesimulation} describes the simulation and the
computational details. In Sec. \ref{sectionphasediagram} we present a
dynamic phase diagram and give an overview of the observed dynamic
phases (\ref{subsectionphasediagram}). These phases are a pinned
vortex glass (\ref{sectionpinnedvortexglass}), different kinds of
turbulent plastic flow (\ref{sectionplasticflow}), a decoupled channel
r\'{e}gime (\ref{subsectiondecoupledchannels}) and coherently moving
structures (\ref{seccoherentlymovingstructure}). In Sec.
\ref{sectiondecoupledchannels} we consider the decoupled channel
r\'{e}gime in detail: we report on the dependence of the spatial
distribution of velocities in the different channels on the pinning
landscape (\ref{chpdynphassampledependenceofmovingtransverseglass}),
we show how topological defects between the channels mediate the phase
slip between channels while preserving the transverse periodicity of
the system (\ref{secslidingmechanisms}), and we give information on
the transverse de-pinning in the decoupled channel r\'{e}gime
(\ref{sectioncriticaltransverseforce}). Finally, we draw our
conclusions in Sec. \ref{sectionconclusions}. Appendix
\ref{secsmoothcutoff} contains technical information on the smooth
 cut-off used.

\section{The simulation}\label{sectionthesimulation}
\subsection{Equation of motion}

We consider a two-dimensional vortex system and model the vortex
motion with overdamped Langevin dynamics. The total force ${\bf F}_i$
acting on vortex $i$ is given by
\begin{equation}
{\bf F}_i= -\eta {\bf v}_i + {\bf F}^{\text{L}} + {\bf F}_i^{\text{vv}} + {\bf F}_i^{\text{vp}} + {\bf F}_i^{\text{therm}} = \bf{0} 
\label{eqnvvinteractionforce}
\end{equation}
where $\eta$ is the Bardeen-Stephen\cite{Bardeen65} viscosity
coefficient, ${\bf v}_i$ the velocity, $ {\bf F}^{\text{L}}$ the
Lorentz force acting equally on all vortices, ${\bf F}_i^{\text{vv}}$
the vortex-vortex interaction, ${\bf F}_i^{\text{vp}}$ the
vortex-pinning interaction, and ${\bf F}_i^{\text{therm}}$ a
stochastic noise term to model temperature.\cite{Chaikin95} The
vortex-vortex interaction force for rigid vortices in thin films and
pancakes in decoupled layers of multi-layer materials experienced by
vortex $i$ at position ${\bf r}_i$ is\cite{Clem91}
\begin{equation} 
{\bf F}_i^{\text{vv}}= \frac{\Phi_0^2 s}{ 2\pi\mu_0\lambda^2} \sum_{j \ne i} \frac{{\bf r}_i-{\bf r}_j}{|{\bf r}_i-{\bf r}_j|^{2}}.
\end{equation} 
The constant $\Phi_0$ is the magnetic flux quantum, $s$ the length of
the vortex, $\mu_0$ the vacuum permeability and $\lambda$ the London
penetration depth. We employ periodic boundary conditions and cut off
the logarithmic vortex-vortex repulsion potential
smoothly.\cite{Fangohr00c} The important feature of this modified
interaction potential is that it does not introduce numerical
artefacts, such as topological defects which can result from using a
na\"{\i}ve cut-off potential. Details can be found in appendix \ref{secsmoothcutoff} and Ref. \onlinecite{Fangohr00c}. The cut-off
distance is $\min(L_x/2,L_y/2)$ where $L_x$ and $L_y$ are the lengths
of the sides \label{txtcutoff} of the rectangular simulation cell. The
lengths $L_x$ and $L_y$ are chosen such that a hexagonal lattice fits
perfectly in the simulation cell.

\begin{figure}
\centerline{\epsfxsize=7cm \epsfbox{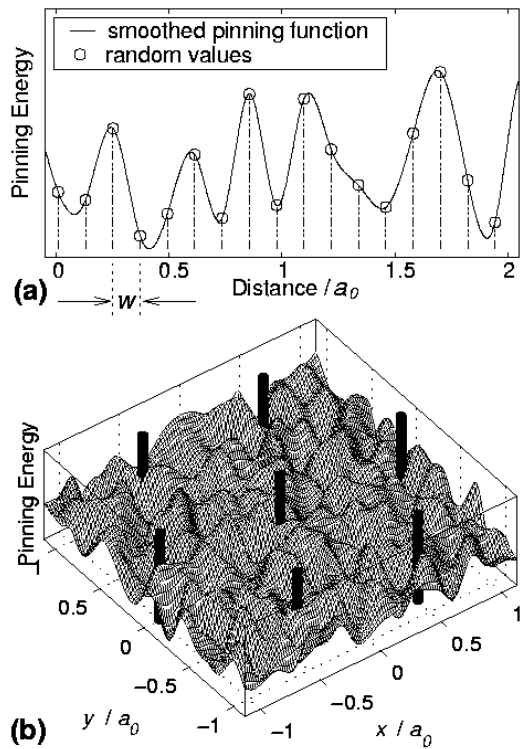}}
\caption{ A sample pinning potential. Distances are
  given in multiples of the vortex lattice spacing, $a_0$. \emph{(a)}
  Demonstration of construction of the pinning potential in one
  dimension: Firstly, we assign random pinning energies at discrete
  sites (shown as open circles) with spacing $w$. Secondly, we
  interpolate between those sites using cubic splines to obtain an
  effectively continous pinning potential. This results in a random
  pinning potential with a short-range correlator
  $\overline{V(r)V(r')}=g(r-r')$ of range $w$. We follow an analogous
  procedure in two dimensions. \emph{(b)} A part of a pinning
  potential as used in the simulations. The seven black cylinders
  indicate vortex lines separated by $a_0$ to demonstrate the length
  scale. \label{figure1}}
\end{figure}

We investigate systems with a magnetic induction of $B=1 \text{ T}$
and a penetration depth of $\lambda = 1400 \text{\AA}$ which yields a
vortex density of $\approx 10/\lambda^2$ representative of typical
cuprate superconductors. The random pinning potential we have employed
varies smoothly on a length scale of $\lambda/25$ which is of the
order of the coherence length $\xi$. This is a representation of
random pinning on the atomic length scale (for example due to oxygen
vacancies or small clusters of oxygen vacancies) since the vortex
cores effectively smooth the pinning potential over a length scale of
the core diameter $2\xi$.  Fig.\ \ref{figure1} (a) demonstrates the
construction of the pinning potential in one dimension. Fig.\ 
\ref{figure1} (b) shows a part of the pinning structure used for the
two dimensional system.  System sizes from 100 to 3000 vortices have
been investigated. We measure lengths in units of
$\lambda=1400\text{\AA}$, and forces in units of the force $f_0$ that
two vortices separated by $\lambda$ experience. We express time in
units of $t_0 =\eta\lambda/f_0 = \eta 2\pi\mu_0\lambda^4/\Phi_0^2 s$
which is in line with other simulations.\cite{Ryu96,Otterlo98}

\subsection{Observables}\label{sectionobservalbes}

To distinguish different dynamic phases we monitor the topological
defect density $\defden$ (defined as the fraction of vortices with
less or more than six nearest neighbors in the Delaunay
triangulation\cite{deBerg97}) and the distribution
$\Gamma(<\!\!v\!\!>)$ of time-averaged velocities
\mbox{$<\!\!v_i\!\!>(t) = |\frac{{\bf {r}}_i(t+t_0)-{\bf
      {r}}_i(t_0)}{t} |$} of individual vortices $i$ over time $t$. We
also observe the structure factor of the system (the Fourier transform
of the vortex positions), a measure for local hexagonal order (using
bond angles $\theta_k$ from the Delaunay triangulation we compute
$\Psi_6=\frac{1}{n_{\text{bond}}} \left|\sum_{k=1}^{n_{\text{bond}}}
  \exp(i6\theta_k)\right|$, where $n_{\text{bond}}$ is the number of
angles in the Delaunay triangulation), the frequency spectrum of the
center of mass velocity, and the paths of motion of vortices
(two-dimensional histogram of vortex positions). We create movies of
time snap-shots of vortex positions to visualize the behavior of the
system.

\section{The dynamic phases}\label{sectiondynamicphases}\label{sectionphasediagram}

\begin{figure}
\epsfxsize=8cm 
\epsfbox{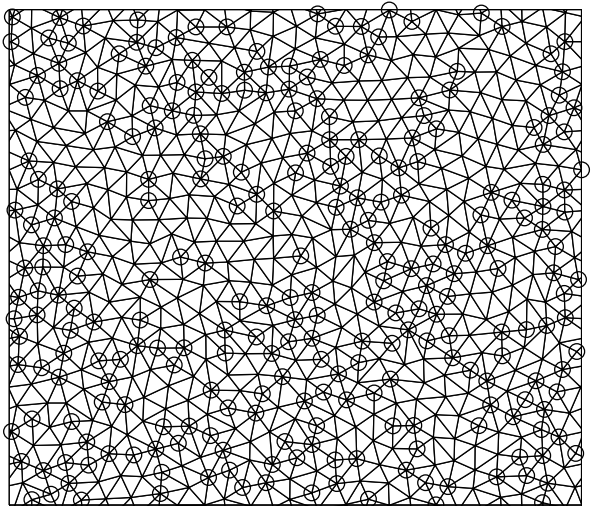}
\caption{
  A pinned vortex glass at pinning strength
  $F^{\text{vp}}_{\text{rms}}=1.2f_0$. The system has been annealed
  from molten to zero-temperature. The figure shows the
  Delaunay\cite{deBerg97} triangulation of vortex positions. Here,
  $\defden = 0.46$, {\it i.e.} 46\% of the 576 vortices are
  topological defects (having more or less than six nearest neighbors)
  and are highlighted by open circles. \label{figure2}}
\end{figure}

Initially, we anneal the vortex system from a molten state to zero
temperature in the presence of the random pinning potential. The
pinning forces are obtained by numerically differentiating the
potential. The root mean square value of the pinning force field is
denoted by $F^{\text{vp}}_{\text{rms}}$. An annealed vortex
configuration is shown in Fig.\ \ref{figure2}. After annealing, a
driving force is applied which is subsequently increased every
$4\cdot10^{4}$ time steps. This yields force-velocity characteristics
which correspond to experimentally obtainable current-voltage
characteristics (at zero-temperature). The driving force ${\bf
  F}^{\text{L}}$ is related to the current density ${\bf j}$ via ${\bf
  F}^{\text{L}}=s {\bf j}\times {\bf \Phi_0} $, and the vortex
velocity to the induced electric field ${\bf E}$ via ${\bf E}={\bf
  B}\times{\bf v}$, where ${\bf B}$ is the magnetic induction and
${\bf v}$ the vortex velocity. We investigate the modes of motion at
different driving forces and pinning strengths using the observables
specified in Sec. \ref{sectionobservalbes}.

\subsection{The phase diagram}\label{secthephasediagram}\label{subsectionphasediagram}

The different observed modes of plastic and elastic motion are
summarized in Tab.~\ref{taboverview} (on page \pageref{taboverview}). The second column of the table
shows the expressions used for each mode of motion and a reference to
the section in which it is described. The third and fourth columns
show the criterion used to identify and distinguish the modes, and the
fifth column gives further observations.

\begin{figure}
\centerline{\epsfxsize=8cm 
\epsfbox{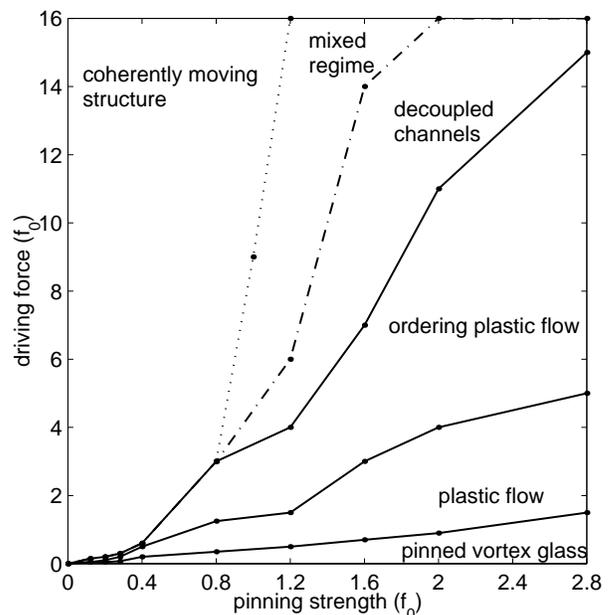}}
\caption{
  Dynamic Phase Diagram of the vortex state as a function of
  disordering pinning strength, $F^{\text{vp}}_{\text{rms}}$, and
  strength of the driving force (at zero temperature).}
\label{figure3}
\end{figure}

We describe now briefly the phase diagram shown in Fig. \ref{figure3}.
For weak pinning ($F^{\text{vp}}_{\text{rms}}\lesssim 0.8f_0$) a
pinned vortex glass (Fig.~\ref{figure2}) undergoes Plastic Flow (PF)
and Ordering Plastic Flow (OPF) for an increasing driving force (Sec.
\ref{sectionplasticflow}). In OPF, in contrast to PF, the density of
topological defects, $\defden$, is lower than the density of the
static system, $\statdef$. We summarize PF and OPF as turbulent
plastic flow because in both modes the motion of vortices is turbulent
rather than laminar, \ie the motion of different vortices is hardly
correlated. This helps us to distinguish between the turbulent
(chaotic) plastic flow of PF and OPF and the (laminar) plastic motion
of vortices in the decoupled channel r\'{e}gime (Sec.
\ref{subsectiondecoupledchannels}, Sec.
\ref{sectiondecoupledchannels}). At high driving forces the vortices
move elastically as a coherently moving structure, \ie every vortex
keeps its nearest neighbors for all times (Sec.
\ref{seccoherentlymovingstructure}).

For the weakest pinning strength of $F^{\text{vp}}_{\text{rms}}\approx
0.04 f_0$ the annealed system is a defect-free Bragg
glass\cite{Giamarchi94} and changes directly from the pinned Bragg
glass to the moving Bragg glass\cite{Giamarchi96} (MBG) without
undergoing plastic motion. However, since the initial configuration is
annealed from random positions, this MBG is, in general, not aligned
with the direction of the driving force, and the pinning is too weak
to reorientate it. This very weak pinning r\'{e}gime has not been
studied in detail in the framework of this investigation.

For stronger pinning ($F^{\text{vp}}_{\text{rms}}\gtrsim 0.8f_0$)
there is an intermediate r\'{e}gime between turbulent plastic flow and
coherently moving structures in which rows of vortices are aligned
with the driving force and vortices move in preferred channels.
However, these channels are decoupled: vortices in different channels
move with different velocities (Sec. \ref{sectiondecoupledchannels}).
The change from the decoupled channel r\'{e}gime to a coherently
moving structure depends on the history of the system: in the mixed
r\'{e}gime both modes of motion can be found depending on whether the
driving force is increased or decreased.

Related numerical work on dynamic phases has been performed by Moon,
Scalettar and Zim\'{a}nyi,\cite{Moon96} Ryu \etal\cite{Ryu96} and
Olson, Reichhardt and Nori.\cite{Olson98} It was found in Ref.
\onlinecite{Moon96} that as the driving force is increased, firstly
the pinned vortex glass exhibits Plastic Flow and finally moves as a
``moving glass'' which is very likely to be the decoupled channel
r\'{e}gime. Ref. \onlinecite{Ryu96} found an elastically moving
structure with topological defects at high driving forces. In contrast
to this work in which we have used logarithmic interactions and have
varied the strength of the pinning forces, in Ref.
\onlinecite{Olson98} the strength of an exponentially decaying
vortex-vortex interaction has been varied in a system with a smaller
vortex density. However, the results can be compared qualitatively,
and Ref. \onlinecite{Olson98} demonstrates similar findings on Plastic
Flow, decoupled channels and coupled channels.

\subsection{Pinned vortex system}\label{sectionpinnedvortexglass}

For sufficiently small driving forces the system is pinned and the
velocity distribution shows a single peak at zero velocity. For
pinning strengths above $\approx 0.04 f_0$ we see the number of
topological defects increasing with pinning strength and no long range
order exists. We thus refer to the pinned system as a vortex glass,
and such a configuration is shown in Fig.\ \ref{figure2}.

\subsection{Turbulent Plastic Flow}\label{sectionplasticflow}

Vortices start moving if the driving force exceeds a critical value.
We distinguish two different kinds of motion which we refer to as
Plastic Flow (PF) and Ordering Plastic Flow (OPF). Both types of
motion show a broad distribution of time-averaged vortex velocities as
shown in the inset in Fig. \ref{figure4}(a), which, for very small
driving forces, has another peak at zero velocity. We call the motion
OPF if the density of topological defects, $\defden$, is below the
defect density, $\statdef$, the system would have if no driving force
was applied. Otherwise we call it PF (Tab. \ref{taboverview}).

We observe PF for driving forces just above the critical de-pinning
force. The topological defect density is higher than for the static
system because some vortices are stationary and others are squeezing
past them. Within the PF r\'{e}gime we find two modes of motion: For
driving forces just above the de-pinning current we find a bimodal
distribution in the time-averaged vortex velocity showing a peak at
zero velocity. Thus, there are some vortices that are permanently
pinned (at least over the simulated time). By contrast, for higher
driving forces, whilst at any one time some vortices may be
stationary, no vortices are permanently pinned. These data confirm
earlier findings of Spencer and Jensen\cite{Spencer97} employing a
simpler model. For clarity, Fig.\ \ref{figure3} does not distinguish
between these two types of PF.

In the OPF r\'{e}gime, where the topological defect density,
$\defden$, is lower than for the static system, we observe that the
instantaneous velocity distribution shows no peak at zero velocity
(\ie all vortices are in motion), whereas this is not the case for
both types of PF described in the last paragraph.  Faleski, Marchetti
and Middleton\cite{Faleski96} used the term crinkle flow to describe
motion of vortices via correlated displacements of patches of
vortices. Our observations of OPF suggest that the definition used
here for OPF ($\defden < \statdef$) is equivalent to the definition of
crinkle flow introduced by Faleski, Marchetti and
Middleton\cite{Faleski96} (absence of a peak at zero velocity in the
instantaneous velocity distribution).

\subsection{Decoupled channels}\label{subsectiondecoupledchannels}
\begin{figure}
\centerline{\epsfxsize=14cm
\epsfbox{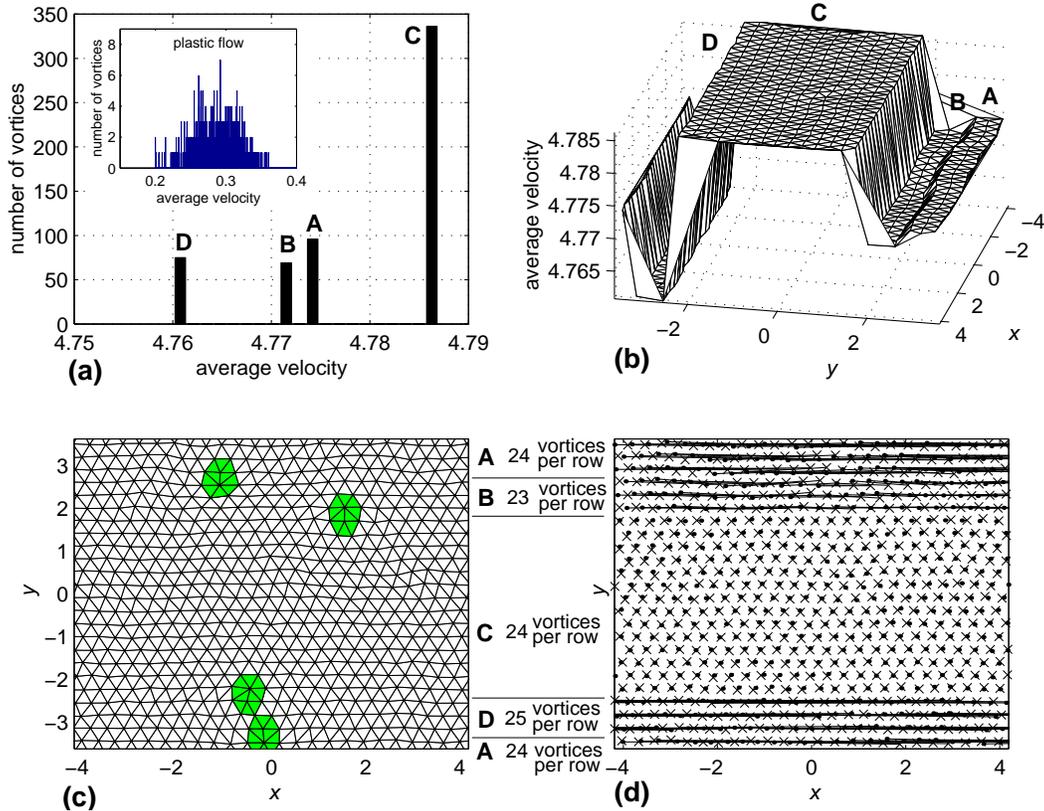}}
\caption{
  The decoupled channel r\'{e}gime. Periodic boundary conditions are
  applied in $x$ and $y$ directions. \emph{(a)} Velocity histogram of
  time-averaged individual vortex velocities $\Gamma(<\!\!v\!\!>)$ for
  a driving force $F^{\text{L}}_x=5.0 f_0$. Inset: same for turbulent
  plastic flow with $F^{\text{L}}_x=0.9 f_0$. \emph{(b)} Delaunay
  triangulation of one time step with the time-averaged individual
  vortex velocity plotted in the third dimension. There are four
  distinct groups labeled A,B,C and D of vortex-channels travelling
  along the $x$-direction with different velocities. \emph{(c)}
  Two-dimensional view of \emph{(b)} and topological defects are
  highlighted. \emph{(d)} Change between initial ($\bullet$) and final
  ($\times$) vortex positions in the frame of reference of one of the
  vortices in group C. \label{figure4} }

\end{figure}

For sufficiently strong pinning and intermediate driving forces (Fig.
\ref{figure3}) we find that vortices arrange in lines orientated along
the direction of the driving force (Fig. \ref{figure4}). These lines
move with different velocities in the direction of the driving force.
This type of motion is described in detail in Sec.
\ref{sectiondecoupledchannels} and is called decoupled channel motion.

\clearpage
\widetext
\parbox{\textwidth}{
\begin{table}
\begin{tabular}{ p{1cm}| p{5cm}| p{2cm} | p{2.2cm}|p{7cm}|}
 & Name &\multicolumn{2}{c|}{Criterion}&Observations  \\ \cline{3-4}
 &      &  $\quad\: \Gamma(<\!\!v\!\!>)$  & $\: \qquad \defden$ & \\ \hline 
Plastic modes & Plastic Flow (PF) and some vortices permanently pinned Sec. \ref{sectionplasticflow}& Broad, and peak at zero & $\defden \ge\statdef$  &  Turbulent flow, system partly pinned  \\ \cline{2-5}
& \raggedright Plastic Flow (PF) and no vortices permanently pinned Sec. \ref{sectionplasticflow} & Broad &$\defden \ge\statdef$ & Turbulent flow, peak at zero in instantaneous velocity distribution (\ie some stationary vortices)
\\ \cline{2-5} \cline{2-5} 
& \raggedright Ordering Plastic Flow (OPF) \newline Sec. \ref{sectionplasticflow} & Broad & $\defden<\statdef$ & Turbulent~flow,~no vortices have zero velocity in instantaneous velocity distribution $\Leftrightarrow$ ``crinkle motion'' (\ie all vortices moving)
\\ \cline{2-5} \cline{2-5} 
& \raggedright  Decoupled channels, \newline Moving Transverse Glass (MTG) \newline Sec. \ref{sectiondecoupledchannels} &  Separated \mbox{$\delta$-peaks} & $0 \! <\! \defden \! \ll \!\statdef$ & Motion in uncoupled channels in direction of driving force, topological defects between channels, critical transverse force 
\\ \hline 
Elastic modes&\raggedright Coherently moving structure without defects, Moving Bragg Glass (MBG), Sec. \ref{seccoherentlymovingstructure}& Single $\delta$-peak & $\defden=0$& Motion in coupled channels in direction of driving force, washboard frequency in noise spectrum, critical transverse force 
\\ \cline{2-5}
& \raggedright Coherently moving structure with defects, Sec. \ref{seccoherentlymovingstructure} & Single $\delta$-peak&$0 \! <\! \defden \! \ll \!\statdef$ & Vortices generally aligned with the direction of the driving force
\\ 
\end{tabular}
\caption{ Overview of observed 
plastic and elastic modes of motion. $\Gamma(<\!\! v \!\!>)$ is the 
distribution of time averaged vortex velocities, and $\defden$ is the 
density of topological defects (Sec. \ref{sectionobservalbes}). The 
topological defect density of the annealed system without any applied 
driving force is $\statdef$. \label{taboverview}
}
\end{table}
}

$\phantom{ }$\\
\narrowtext
\clearpage

\clearpage
\widetext
$ $\vspace{1cm}
\begin{figure}
\epsfxsize=14cm
\centerline{\epsfbox{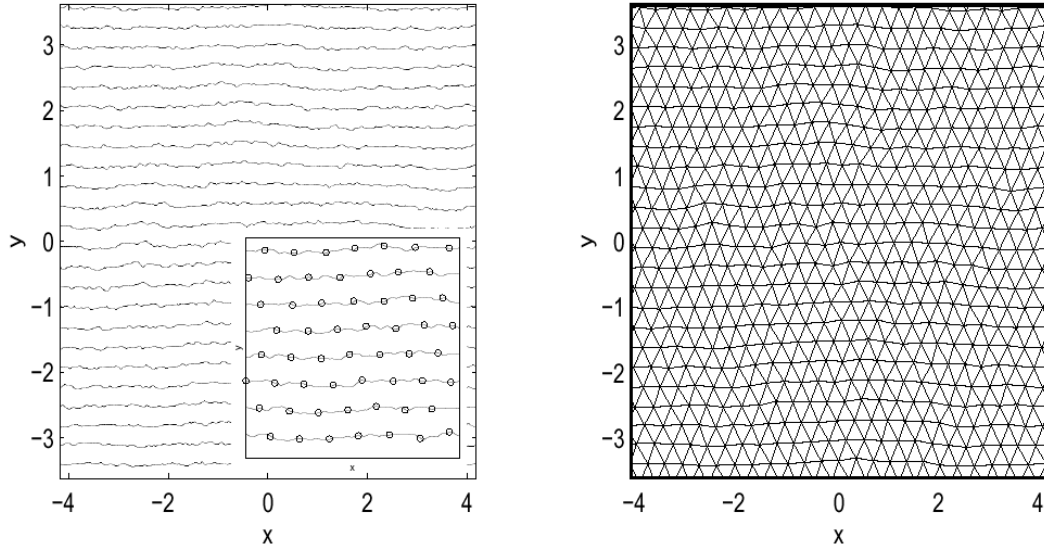}}
\caption{
  The Moving Bragg Glass. \emph{Left:} Histogram of vortex positions.
  The driving force is acting from left to right along the
  $x$-direction, and vortices move in rough channels, like beads on a
  string. The inset shows a slightly enlarged version of the channels
  and positions of vortices for one time step are shown as circles.
  \emph{Right:} Delaunay configuration of one snap shot of the same
  system. Although the channels in the left plot are rough, there are
  no topological defects in the Moving Bragg Glass.
 \label{figure5} }
\vspace{-0.2cm}
\end{figure}

\vspace{2cm}
\begin{figure}
\epsfxsize=6cm
\centerline{\epsfbox{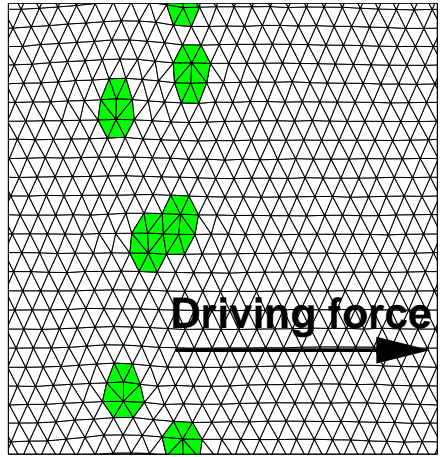}}
\caption{
  A coherently moving structure with 6 dislocations (12 topological
  defects).
\label{figure6}}
\end{figure}

\narrowtext
\clearpage

\subsection{Coherently moving structure}\label{seccoherentlymovingstructure}\label{txtmovingbraggglass}

We observe two different kinds of coherently moving structures: (i)
either a Moving Bragg Glass\cite{Giamarchi96} (MBG) which
is free of topological defects (Fig. \ref{figure5}), or (ii) a
hexagonal system similarly aligned with the direction of the driving
force but with a few dislocations (Fig.\ \ref{figure6}). Both
configurations move elastically, \ie each vortex keeps its nearest
neighbors for all times.

For strong pinning ($F^{\text{vp}}_{\text{rms}}\gtrsim 0.8f_0$) and
increasing driving force the transition from the decoupled channel
r\'{e}gime to a coherently moving structure results in a MBG if the
groups of coupled channels have the same vortex line density (see
Sec.~\ref{secslidingmechanisms}). If the groups of coupled channels
have different line densities then the dislocations between them are
frozen into the coherently moving structure. For weak pinning
($F^{\text{vp}}_{\text{rms}}\lesssim 0.8f_0$) the vortices do not move
in decoupled channels for intermediate driving forces, and the system
changes directly from Ordering Plastic Flow (OPF) to a coherently
moving structure. Again, elastically moving systems with and without
topological defects are observed. Our data from simulating
current-voltage characteristics with increasing driving force suggest
that the configuration at high driving forces is usually a hexagonal
system aligned with the driving force with a few dislocations.
However, a MBG configuration is occasionally observed.

The coherently moving structures we observe are always aligned with
the direction of the driving force for pinning strengths $\gtrsim
0.8f_0$. For smaller pinning strengths, configurations develop
occasionally which are not aligned with the driving force. It has been
argued\cite{Schmid73,Muellers95,Balents98,LeDoussal98} that this
alignment minimizes power dissipation, and our results are in
agreement with other numerical investigations\cite{Moon96,Ryu96} in
which the high-velocity configurations are generally aligned with the
driving force.

For a MBG we find peaks at multiples of the washboard frequency
$\omega_0=2\pi<\!\!v_{\text{cm}}\!\!>/a_0$ in the Fourier spectrum of
the center-of-mass velocity $v_{\text{cm}}(t)$ of the system, where
$<>$ denotes a time-average and $a_0$ is the lattice constant of the
vortex lattice. Whereas this temporal periodicity is not existent for
the velocity of an individual vortex, we also find it in the energy of
the system. Clear peaks in the Fourier spectrum can be observed up to
frequencies of $\approx 100 \omega_0$. For a single particle the
washboard frequency is observable if it slides through a periodic
potential. Here, we have a random potential, but a periodic system.
One thus finds the washboard frequency in observables that depend on
all vortices, such as the center-of-mass velocity or the energy, but
not for individual vortices. The washboard frequency has been found
experimentally in AC\cite{Bhattacharya91,Harris95,Harris96} and
DC\cite{Togawa00} measurements, and numerically\cite{Olson98} in a
similar r\'{e}gime.

In summary, we observe occasionally a MBG at high driving forces, but
most of the final configurations are hexagonal systems aligned with
the driving force with a few dislocations pairs. However, it could
well be that finite temperatures or larger systems would favor the
creation of a MBG at high driving forces: as yet, it is not clear what
is the ``dynamic ground state'' of these systems. Our data cannot be
used to decide whether a MBG exists in two-dimensions or whether the
MTG is the only stable phase,\cite{Balents98,LeDoussal98} since for a
system of a given size if the velocity is sufficiently large then all
channels couple and appear to be a MBG. Another open question is
whether periodic boundary conditions can favor a re-ordering of a
disordered vortex system.\cite{Vinokur00} The exploration of these
questions is computationally expensive though new methods for
evaluating interactions in the system may make this
feasible.\cite{Fangohr00c,Cox00}

\section{Decoupled channels}\label{sectiondecoupledchannels}

This section describes a plastic mode of motion which, due to its
quite different properties, is separated from the Sec.
\ref{sectionplasticflow} on turbulent plastic flow. As visible in the
dynamic phase diagram in Fig.~\ref{figure3} the decoupled channels are
only observed for sufficiently strong pinning.

Increasing the driving force (which acts in the $x$-direction) from
the turbulent plastic flow r\'{e}gime further, transforms the
time-averaged velocity distribution from a broad peak as observed for
turbulent plastic flow to several clearly distinct peaks as shown in
Fig.\ \ref{figure4}\ (a). In \ref{figure4} (b) four different velocity
levels are visible, each of these corresponding to one peak in the
velocity histogram. Thus, vortices move in four groups of coupled
channels and, within a group, all channels travel with a constant
velocity in the direction of the driving force. Plot \ref{figure4} (c)
shows the lattice structure of (b) in a two-dimensional projection.
Vortices with more or less than six nearest neighbors are highlighted
by a grey shade. We see that the groups of coupled channels are
separated from each other by one 5-7 dislocation (a pair of vortices
one having 5 and the other 7 nearest neighbors).

Plot \ref{figure4} (d) shows the initial ($\bullet$) and final
($\times$) positions of vortices in the frame of reference of one of
the vortices in group C, and the initial and final positions are
connected by a straight line, demonstrating that vortices never change
the channels in which they move. This is a particularly interesting
point since the moving glass theory\cite{LeDoussal98}
assumes that the topological defects (which are hard to treat
analytically) between groups of coupled channel do not destroy
transverse periodicity. Thus, the observation that these topological
defects do not introduce chaotic motion of vortices and that the
defects just decouple the different channels supports the theory of
Giamarchi and Le\ Doussal.\cite{LeDoussal98}

A series of runs shows that generally for larger driving forces the
number of channels (and thus the size of each group) which are coupled
and move with the same velocity increases, until the system shows
elastic motion. For decreasing driving forces the number of coupled
channels decreases until each group of coupled channels exists of only
one or two channels. For even smaller driving force, the systems
exhibit turbulent plastic flow.

For the mixed r\'{e}gime shown in Fig.\ \ref{figure3} we find that the
mode of motion depends on the history of the system: Increasing the
driving force for a system in the decoupled channel r\'{e}gime into
the mixed r\'{e}gime results in motion in decoupled channels. On the
other hand, reducing the driving force for a Moving Bragg Glass (MBG)
into the mixed r\'{e}gime, yields elastic motion. For driving forces
above the dotted line in Fig.\ \ref{figure3} all systems show elastic
motion and below the dash-dotted line all systems show smectic motion.
The data suggest that both the MBG and the decoupled channels are
metastable steady states which are separated by an energy barrier. In
future we will explore whether finite temperatures are able to
overcome this barrier.

\begin{figure}
\epsfxsize=7cm
\epsfbox{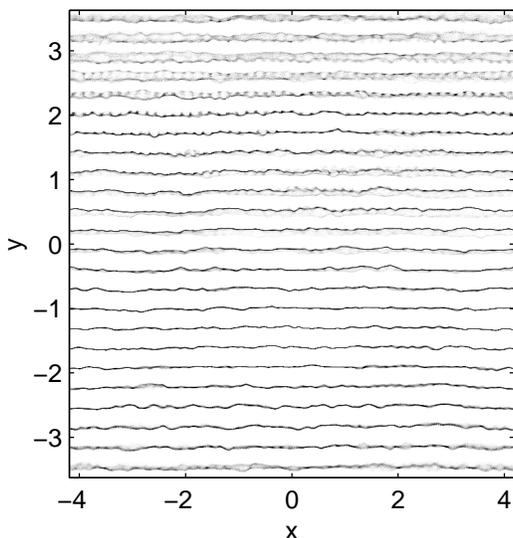}
\caption{
  This plot shows for the decoupled channel r\'{e}gime in which areas
  of the simulation vortices prefer to travel. On a 700$\times$700
  cell grid, a two-dimensional histogram of vortex positions has been
  created. The darker a cell, the more vortices have been counted
  within that cell over the duration of the simulation.
\label{figure7} }
\end{figure}

Fig.~\ref{figure7} shows an accumulation of vortex positions using a
grid of 700$\times$700 cells. It demonstrates that the channels in
which vortices move are not strictly static but slightly broadened
(see, for example, $y\approx 3$), although vortices never change
channels. Further analysis in Sec.~\ref{secslidingmechanisms} shows
that the 5-7~dislocations highlighted in Fig.\ \ref{figure4}\ (c) move
with time in the $x$-direction parallel to the driving force.
Presumably this requires slight corrections of the static channels,
which results in their blurred form visible in Fig.\ \ref{figure7}.
This is supported by results given in Fig.~\ref{figure5} which show
that for the MBG in the absence of dislocations the resulting channels
are strictly static (and not blurred). This may indicate that the
theoretical model\cite{LeDoussal98} predicting strictly
static channels for the MTG may be too simple.

\begin{figure}
\epsfxsize=7cm
\epsfbox{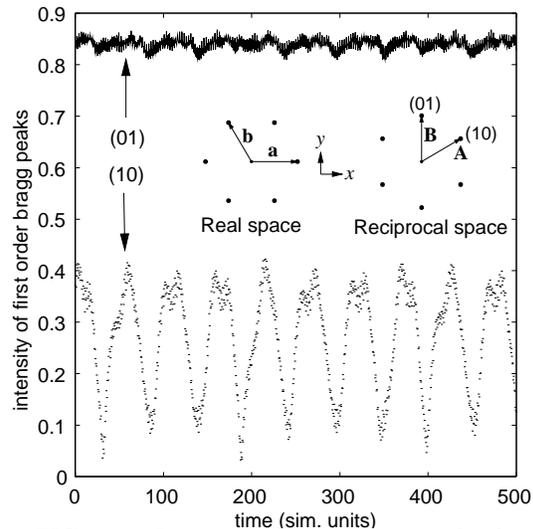}
\caption{
  Intensity of two Bragg peaks for the Moving Transverse Glass shown
  in Fig. \ref{figure4}. The (01) peak is relative constant at a value
  of approximately $0.85$ and it measures the order along the
  $y$-direction. In contrast, the peak (10) measures also order along
  the $x$-direction, and oscillates strongly. The variations are due
  to the different groups of coupled channels sliding past each other.
  The deviation of the (01) peak from 1.0 (as for a perfect lattice)
  is due to the roughness of the channels. The inset shows the lattice
  vectors used to label the peaks. \label{figure8} }
\end{figure}

Fig.~\ref{figure8} shows the square modulus of the structure factor
for two ${\bf k}$-vectors for the MTG. The large value of the
(01)-peak indicates the transverse order of the system. The small and
oscillating (10) peak shows the strongly reduced order in the
$x$-direction. A time Fourier transform of the (10) signal reveals the
frequencies with which the different groups overtake each other by one
lattice spacing. This is exactly what is expected for a MTG in a
finite system and is in agreement with the theoretical
prediction\cite{Balents97,Balents98,Scheidl98b,LeDoussal98} for the smectic
r\'{e}gime that any peaks in the structure factor with a non-zero
$x$-component should vanish in an infinitely large system.

The data shown are in agreement with the theoretically predicted
moving smectic\cite{Balents97,Balents98} which is also called
Moving Transverse Glass\cite{LeDoussal98} (MTG), and decoupled channel
r\'{e}gime.\cite{Scheidl98b} A MTG has previously been identified by a
numerical study of Olson, Reichhardt and Nori,\cite{Olson98} and the
Delaunay triangulation of a snap-shot of their system looks
qualitatively like Fig.~\ref{figure4}~(c). 
Kolton, Dom\'{\i}nguez and Gr{\o}nbech-Jensen 
numerically found smectic states,\cite{Kolton99aandKolton00aandKolton01a} and 
earlier the numerical
studies of Moon, Scalettar and T. Zim\'{a}nyi\cite{Moon96} on moving
vortex systems suggested the possibility of phase slips of different
channels. Further new results on the MTG are presented in the next two
sections concerning the spatial distribution of vortex channel
velocities depending on the pinning landscape
(Sec.~\ref{chpdynphassampledependenceofmovingtransverseglass}), the
mechanism of uncoupled channels sliding past each other
(Sec.~\ref{secslidingmechanisms}), and the transverse de-pinning
(Sec.~\ref{sectioncriticaltransverseforce}).

\subsection{Dependence of the spatial velocity distribution on pinning landscape}\label{chpdynphassampledependenceofmovingtransverseglass}

We find a correlation between the particular pinning potential
employed (representing details of the microstructure causing vortex
pinning in the material) and the positions and velocities of the
different groups of coupled channels. It should be noted that,
although for the data shown in Fig.\ \ref{figure4}\ (b) the vortices
with the maximum velocity are located in the central region of the
sample ($y\approx 0 \pm 1.5$), this is not an edge effect: for other
samples the maximum is located at different $y$-positions. Remarkably,
in both the decoupled channel r\'{e}gime (Fig.~\ref{figure4}) and the
turbulent plastic flow r\'{e}gime (not shown here) the fastest flow is
located in the same part of the simulated material.

It has been suggested\cite{Bhattacharya99} that the different
velocities observed in the decoupled channel r\'{e}gime and shown in
Fig.~\ref{figure4} may be related to the experimentally observed
fingerprint effect.\cite{Higgins96} In fact, it seems that for both
the turbulent plastic flow and the decoupled channel r\'{e}gime the
same areas of the pinning potential allow for better (or worse)
pinning. This is not obvious since in the turbulent plastic flow
r\'{e}gime vortices flow more or less individually along highly
tortuous paths whereas in the decoupled channel r\'{e}gime they move
in a much more correlated way.

\subsection{Channel sliding mechanisms}\label{secslidingmechanisms}

\begin{figure}
\epsfxsize=7cm
\epsfbox{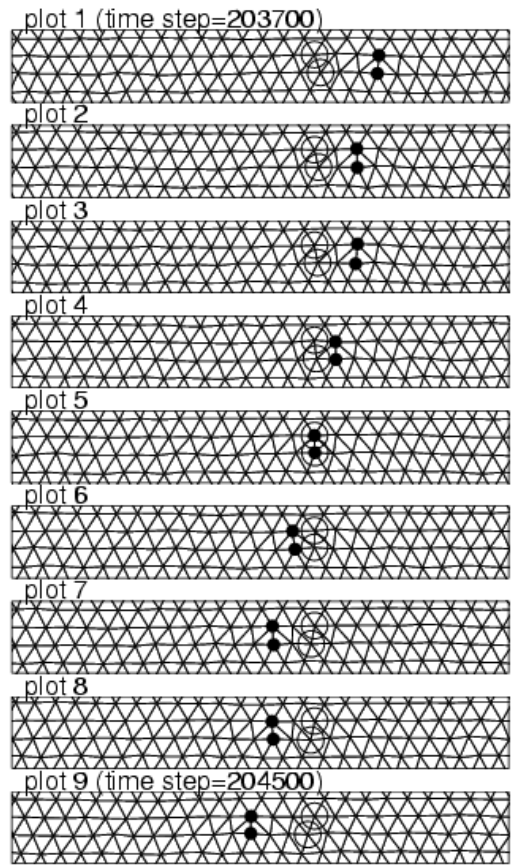}
\caption{
  Dislocation mediated phase slip in decoupled channel r\'{e}gime for
  different longitudinal vortex densities showing snap shots of 9 time
  steps. The open circles in each plot mark two fixed vortices, and
  the upper one defines the frame of reference. The Lorentz force acts
  from left to right. The upper two lines of vortices visible in each
  plot belong to group C in Fig.\ \ref{figure4} and the lower to lines
  to group D. Thus, in the relative frame of reference, the lower two
  lines move to the left. The two black filled circles indicate
  topological defects in each snap-shot and represent a dislocation.
  These mediate the phase slip while moving to the left as can be seen
  by comparing the open circles in plot 1 and 9.
\label{figure9}\vspace{0.35cm}}
\end{figure}

Fig.\ \ref{figure4} shows that the number of vortices per line of
vortices (the line density) differs from group to group by 1, and
exactly one dislocation between the groups is required to accommodate
this difference (\ref{figure4} c). In Fig.~\ref{figure9} a time series
of Delaunay\cite{deBerg97} triangulations of snap-shots of a part of
Fig. \ref{figure4} is shown, demonstrating how a moving dislocation
allows group C to move faster than group D.

\begin{figure}
\epsfxsize=\textwidth
\centerline{\epsfbox{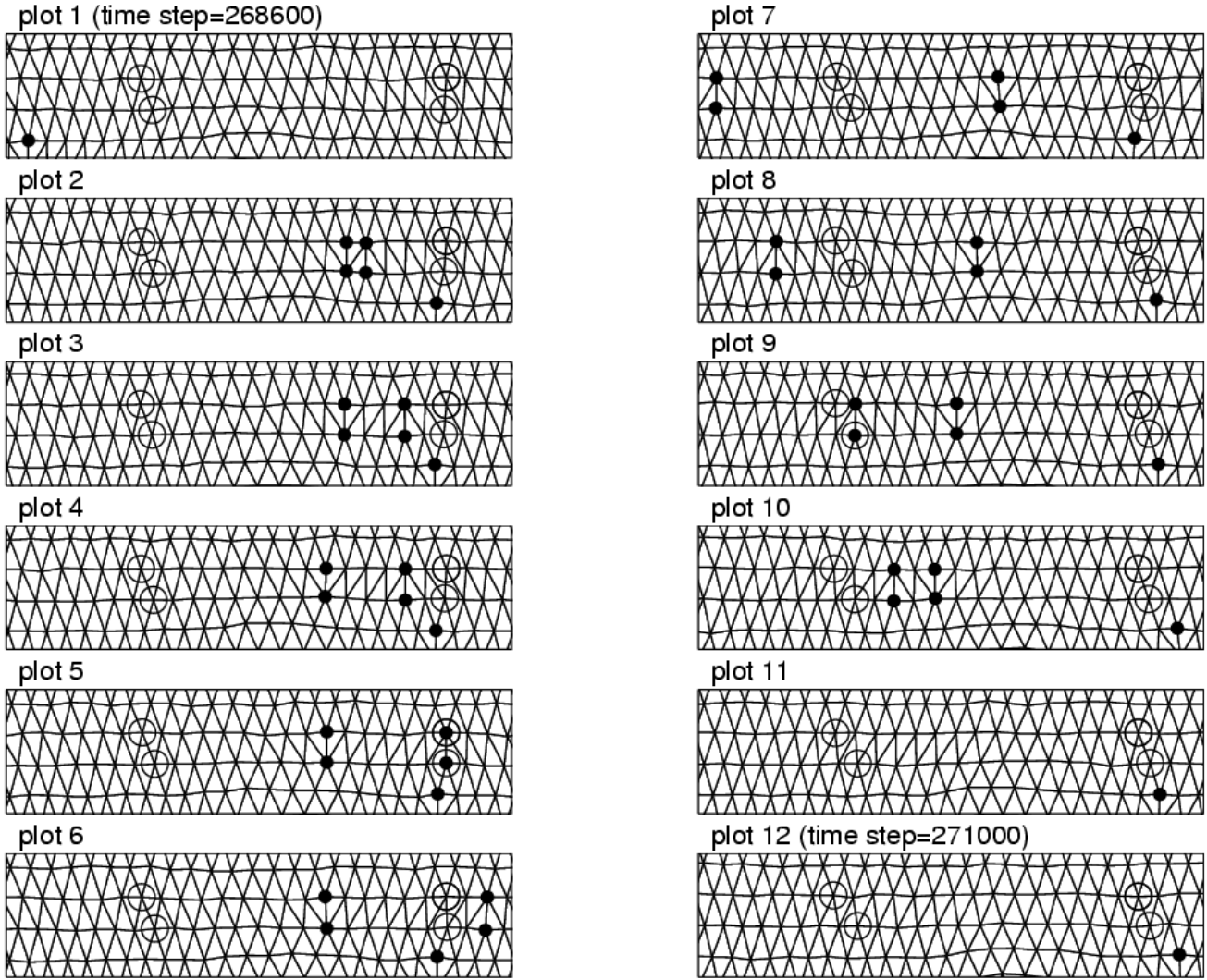}}
\caption{
  Dislocation mediated phase slip in decoupled channel r\'{e}gime for
  equal longitudinal vortex densities. The open circles mark four
  fixed vortices and black filled circles show topological defects.
  Here, in the relative frame of reference, the lower two lines move
  to the right. The phase slip is realized by two dislocations
  consisting of a 5-7 and a 7-5-disclination pair (counting first the
  nearest neighbors for the upper disclination). These emerge from a
  pair-anti-pair creation process in plot~2. The 5-7-dislocation on
  the left moves to the left and the 7-5-dislocation to the right-hand
  side. In this process they allow the lower lines to pass a lattice
  spacing to the right. Between plot~6 and~7 the 7-5-dislocation
  leaves the simulation cell at the right-hand side and enters it
  again on the left-hand side. Finally, in plot~10, the two
  dislocations meet again and annihilate each other. The topological
  defect in the lowest row is not important here.
 \label{figure10}}
\end{figure}

Fig.~\ref{figure10} shows this process for two neighboring channels
having the same longitudinal vortex density. Again the phase slip is
mediated by dislocations which travel along the channel. However,
since initially there are no dislocations between the channels a
dislocation pair is created. These dislocations travel away from each
other, allowing the phase slip between the upper and lower two lines.
When the dislocations meet again (due to periodic boundary conditions)
they annihilate.

Only the two mechanisms shown in Fig.\ \ref{figure9} and
\ref{figure10} have been observed. The situation with the same
longitudinal vortex density has been observed less frequently which
may indicate that this is energetically more expensive. However, in
macroscopic systems the two mechanisms described are less
distinguishable and may coexist: the local vortex line density differs
around each of the dislocations in Fig.\ \ref{figure10}. Thus, the
process shown in Fig.~\ref{figure9} may just be a more detailed study
of the phase slipping process in Fig.~\ref{figure10} for each of the
dislocations.

From a figure in the work of Olson, Reichhardt and Nori\cite{Olson98}
we identify varying longitudinal vortex densities and Burgers
vectors\cite{Ashcroft76} parallel to the driving force, both in
agreement with our results.

In conclusion, the detailed mechanism of decoupled channels moving
past each other has been identified for the first time. The phase slip
is mediated by disclination pairs which either exist between separate
groups of coupled channels with locally different line densities of
vortices, or the disclination pairs are created dynamically and in
pairs when sufficient shear stress has built up. These results may
help in finding a starting point for a theoretical description of the
dynamics of dislocations, such as a
Kosterlitz-Thouless\cite{Kosterlitz73} theory for non-equilibrium
systems.

\clearpage
\widetext
$ $
\vspace{4cm}
\narrowtext
\clearpage

\subsection{Transverse de-pinning}\label{sectioncriticaltransverseforce}
Following the theory of Giamarchi and Le Doussal,\cite{Giamarchi96} at
zero temperature, a non-analytical response of the vortex system to a
small transverse force is expected for the moving glass, \ie for the
MBG and the MTG. In agreement with our results in both the MBG and the
MTG the existence of such transverse barriers have been observed in
simulations.\cite{Moon96,Ryu96,Olson00b,Fangohr00a} The transverse
de-pinning of the MBG has recently been described\cite{Olson00b} and
here we report on the transverse de-pinning of the MTG. We have found
that the transverse de-pinning of a MTG may happen in two ways:

Mechanism A: At a certain strength of the transverse force (below the
transverse de-pinning force), some vortices change the rows in which
they have moved so far, such that after this first step the different
groups of coupled channels A, B, C and D (in Fig.~\ref{figure4}) have
the same number of vortices in each row. Then, all these rows move
with the same velocity, \ie the system has changed from a MTG to a
MBG. This MBG does not (yet) move in the transverse direction. Only
when the transverse driving force is increased even further, the
system de-pins in the transverse direction, and moves elastically in
the longitudinal and the transverse direction as described in Ref.
\onlinecite{Olson00b}.

Mechanism B: At the transverse de-pinning force the system rearranges
plastically such that the rows of vortices (which are aligned with the
$x$-axis in Fig.~\ref{figure4}) become orientated with an angle to the
$x$-axis after the change. The new direction of the rows is not the
same as the direction of the total driving force (adding the small
transverse force to the main driving force along the $x$-axis).

We have found that the critical transverse force is higher for
mechanism A. Our early investigations have shown that the transverse
de-pinning of the MTG is an intricate matter and further studies are
required to reveal under which circumstances mechanism A or B appears.

\section{Conclusions}\label{sectionconclusions}

We have modelled the dynamics of vortices in two dimensions using
overdamped Langevin dynamics with a logarithmic vortex-vortex
interaction potential which includes a modified
cut-off\cite{Fangohr00c} to avoid introducing numerical artefacts into
the simulation. We have computed a dynamic phase diagram as a function
of pinning strength and driving force. We find pinned vortex systems,
different kinds of turbulent plastic flow, and for large driving
forces motion of vortices in rough channels along the direction of the
driving force. 
Depending on pinning strength and driving force the
motion in different channels can either be coupled or decoupled. These
phases can be identified with the predicted
Moving Bragg Glass\cite{Giamarchi96} (MBG) and the Moving Transverse Glass \cite{Balents97} (MTG) as described in recent theoretical models.\cite{Giamarchi96,Balents97,Balents98,Scheidl98b,LeDoussal98}

We have studied the MTG in detail and report on the dependence of the
vortex channel velocities on the pinning landscape. We have identified
how topological defects mediate the phase slip between channels moving
with different velocities, and we have shown that vortices never
change the channels in which they are moving, \ie the dislocations in
the system do not produce chaotic motion of vortices, thus preserving
transverse periodicity. Together with the observed critical transverse
force for the MTG and the MBG in these simulations, our findings
strongly support the moving glass theory of Giamarchi and Le
Doussal\cite{LeDoussal98} which assumes that the
dislocations in the MTG do not introduce additional effects that may
destroy transverse periodicity (and thus the critical transverse
force) in the MTG.

Our findings may also help in finding an extension to the
Kosterlitz-Thouless\cite{Kosterlitz73} theory for non-equilibrium
systems.

\paragraph*{Acknowledgments} We thank P. Le Doussal, A. R. Price and S. Gordeev for helpful discussions. We acknowledge financial support from DAAD and EPSRC.

\appendix

\section{Smooth cut-off}\label{secsmoothcutoff}
We employ a smooth cut-off for the vortex-vortex interaction following
the ideas described in Ref. \onlinecite{Fangohr00c}. Here we give
details on the particular interpolating function we have chosen.

The vortex-vortex interaction has to be cut off for distances greater
than a cut-off distance, $b$. Assume the interaction force is given by
$f(r)$. For short-ranged interactions it is sufficient to use an
interaction $\hat{f}(r)$ which is $f(r)$ for $r\le b$ and zero otherwise:
$$ \hat{f}(r) = \left\{\begin{array}{ccl} f(r) &:& r \le b \\
0 &:& r > b. \end{array}\right. $$

\begin{figure}
\vspace{2cm}
\epsfxsize=9.5cm
\centerline{\epsfbox{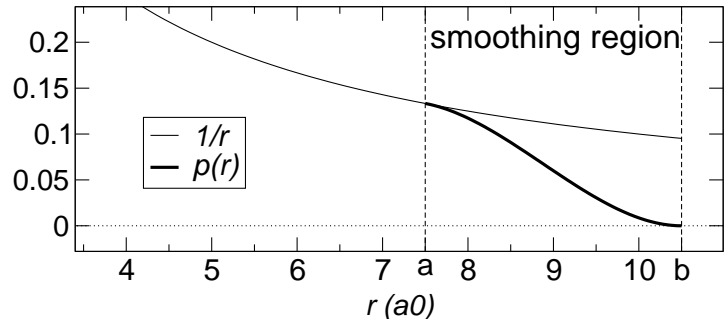}}
\caption{Demonstrating the shape of the interpolating polynomial $p(r)$ (thick line) which smoothly reduces the vortex-vortex interaction force $f(r)$ to zero. For clarity we have chosen $f(r)=1/r$. The interpolation starts at the fading distance $a=7.5a_0$ and reduces the interaction force to zero at the cut-off distance $b=10.5a_0$, where $a_0$ is the average vortex lattice spacing. See text for details.
 \label{figure11}}
\end{figure}

For long-ranged forces this approach results in artificial
configurations.\cite{Fangohr00c} However, those problems can be
overcome by reducing $f(r)$ smoothly to zero near the cut-off distance
$b$. One needs to introduce another distance, $a$, and a polynomial
$p(r)$, such that $a<b$ and that $p(r)$ interpolates
between $f(a)$ at $a$ and and zero at $b$:
$$ \hat{f}(r) = \left\{\begin{array}{ccl} f(r) &:& r \le a \\
                                    p(r) &:& a < r \le b\\
                                    0 &:& r > b.
                 \end{array}\right.$$
It is required\cite{Fangohr00c} that $\hat{f}(r)$ shows ${\mathcal{C}}^1$ continuity at $a$ and $b$, and its derivative at $b$ to be zero:
\begin{mathletters}
\begin{eqnarray}
f(a) &=& p(a), \label{eqn1beg}\\
  p(b) &=& 0, \\
\left.\frac{\D f}{\D r}\right|_{r=a} &=& \left.\frac{\D p}{\D r}\right|_{r=a}, \\
  \left.\frac{\D p}{\D r}\right|_{r=b} &=& 0.  \label{eqn1end}
\end{eqnarray}
\end{mathletters}
\clearpage
\widetext
We have used a third order polynomial $$p(x) = \sum\limits_{i=0}^{3}
c_i x^i = c_3 x^3 + c_2 x^2 + c_1 x + c_0 $$
and the coefficients
$c_i$ are completely determined by equations (\ref{eqn1beg}) to
(\ref{eqn1end}). Writing $f'(r)$ for $\frac{\D f}{\D r}(r)$ one finds 
$$
\left(\begin{array}{c}
c_3 \\ c_2 \\ c_1 \\ c_0 
\end{array}\right)
= \frac{1}{(a-b)^3} \left(\begin {array}{c} f'(a)a-f'(a)b-2\,f(a)
\\\noalign{\medskip}-f'(a){a}^{2}+3\,f(a)(a+b)-af'(a)b+2\,f'(a){b}^{2}\\\noalign{\medskip}\left (2\,f'(a){a}^
{2}-af'(a)b-6\,f(a)a-f'(a){b}^{2}\right )b
\\\noalign{\medskip}-f'(a){b}^{2}{a}^{2}+3\,f(a)a{b}^{2}+f'(a)a{b}^{3}-{b}^{3}f(a)
\end {array}\right ).
$$
The cut-off distance $b$ is determined by geometrical constraints
(see Sec. \ref{txtcutoff}). We follow Ref. \onlinecite{Fangohr00c}
and choose the distance $b-a$ over which the interaction is reduced to zero 
to be three lattice spacings, so
that $a=b-3a_0$. In this work $f(r) \propto 1/r$. Fig. \ref{figure11}
shows a schematic plot of the smooth cut-off and the interpolating
polynomial.

To compute the potential energy of the system it is
required to integrate $-p(x)$ to represent the smoothed interaction
potential for $a < r < b$. The integration constant is determined by requiring
continuity of the interaction potential at $r=a$.

\clearpage
\narrowtext

\end{document}